\documentclass[conference]{IEEEtran}

\usepackage{cite}
\usepackage{amsmath,amssymb,amsfonts}
\usepackage{graphicx}
\usepackage{textcomp}
\usepackage{xcolor}

\usepackage{booktabs}

\usepackage{url}

\usepackage[hidelinks]{hyperref}

\title{Narrative-Centered Emotional Reflection: An Early Prototype for AI-Supported Emotional Self-Reflection}

\author{
Shou-Tzu Han \\
Independent Researcher \\
contact@reflexionai.dev
}
\begin{document}
\maketitle

\begin{abstract}
Reflexion is an AI-powered prototype designed to explore structured emotional self-reflection. By integrating emotion detection, layered reflective prompting, and metaphorical storytelling generation, Reflexion was intended to support users in autonomous emotional exploration beyond basic sentiment categorization. Grounded primarily in expressive writing, cognitive restructuring, and self-determination theory, the system was designed to organize reflection as a progressive pathway from surface-level emotional recognition toward value-aligned action planning. Its final action-planning layer is additionally informed by broader questions of agency and empowerment, which remain future directions rather than fully implemented mechanisms in the current prototype. Informal design feedback indicated that some reviewers found the layered interaction model understandable and potentially useful; no empirical efficacy claims are made. As an early prototype, Reflexion documents one direction in theory-informed affective computing.
\end{abstract}

\begin{IEEEkeywords}
Affective computing, emotional literacy, human-centered AI, narrative psychology, reflective technologies
\end{IEEEkeywords}

\section{Introduction}

Self-reflection on emotional experiences has been studied as one pathway through which individuals can organize, process, and make meaning of emotionally significant experiences. Prior work on expressive writing suggests that structured emotional expression can support reflection and meaning-making in some contexts, although reported effects and mechanisms vary \cite{pennebaker1997opening}. Despite a growing global emphasis on mental health, particularly among youth and young adults, technological tools that support nuanced, autonomous emotional exploration appear comparatively underexplored relative to categorical mood tracking and prescriptive wellness applications.

Existing AI-based emotional tools, such as chatbots and mood trackers, often simplify emotional complexity into discrete categories or deliver overly generalized wellness suggestions. These approaches risk flattening emotional experiences rather than deepening users' understanding, and frequently fail to provide the narrative structure or reflective scaffolding necessary for meaningful emotional development \cite{fitzpatrick2017delivering}.

Reflexion addresses this gap by introducing an AI-powered platform that integrates emotion detection, layered reflective prompting, and metaphorical storytelling generation to enable scalable, autonomous emotional self-reflection. Unlike diagnostic or prescriptive systems, Reflexion invites users into an iterative, user-led journey of emotional sense-making, drawing from theories of expressive writing \cite{pennebaker1997opening}, cognitive restructuring \cite{beck2021cognitive}, and narrative identity development \cite{bruner1990acts}.

In this paper, the name Reflexion refers to an affective self-reflection prototype for narrative-centered emotional exploration. This use is distinct from prior work using ``Reflexion'' to describe verbal reinforcement learning for language agents \cite{shinn2023reflexion}; the present system focuses on user-facing emotional reflection, metaphorical narrative generation, and structured self-inquiry rather than agent learning or task-solving.

Recognizing that emotional growth is nonlinear and individualized, Reflexion offers multiple structured reflection pathways, enabling users to progress from surface-level emotional recognition toward deeper cognitive reframing and value-driven action planning. Its layered architecture is grounded primarily in three established psychological frameworks: expressive writing \cite{pennebaker1997opening}, Cognitive Behavioral Therapy \cite{beck2021cognitive}, and Self-Determination Theory \cite{ryan2000self}. The final layer's emphasis on value-aligned action additionally draws loose inspiration from critical consciousness development \cite{freire1970pedagogy}; we treat this connection as aspirational rather than implemented, and return to it as a direction for future work (Section~\ref{sec:limitations}).

Through this theoretical integration, Reflexion explores one possible design direction for AI-mediated emotional self-reflection, with attention to self-awareness, emotional articulation, and value-aligned sense-making.

Our key contributions are as follows:
\begin{itemize}
\item We introduce Reflexion, a prototype AI-driven platform that integrates emotion detection, layered journaling prompts, and metaphorical storytelling to explore scalable, autonomous emotional reflection.
\item We propose a multi-layered reflection architecture informed by psychological and educational theories, designed to scaffold users' emotional development in a progressive, self-directed manner.
\item We present informal design feedback from preliminary prototype reviews, describing how reviewers perceived the layered interaction model, metaphorical narratives, and autonomy-supportive design.
\item We critically examine system limitations, including cultural specificity and prototype constraints, and outline directions for future work, including empirical validation, backend scaling, and broader evaluation in educational, self-reflection, and personal development contexts.
\end{itemize}

\section{Related Work}

Research at the intersection of affective computing, narrative psychology, and emotional literacy has increasingly emphasized the importance of integrating emotional signals into human-computer interaction systems. Affective computing, pioneered by Picard \cite{picard1997affective}, laid the foundation for recognizing, interpreting, and responding to human emotions computationally. Subsequent advancements have largely focused on emotion detection, sentiment analysis, and the development of empathetic dialogue agents \cite{calvo2014positive}.

In parallel, narrative psychology has highlighted the central role of story construction in human meaning-making. Bruner \cite{bruner1990acts} argued that narratives serve as essential mechanisms through which individuals organize identity and process emotional experiences. Complementing this perspective, educational frameworks such as CASEL \cite{casel2020framework} stress the importance of cultivating emotional awareness, regulation, and expression, particularly within developmental and educational contexts.

Several AI-driven platforms have emerged to support emotional well-being. Woebot \cite{fitzpatrick2017delivering} and Wysa \cite{inkster2018empathy} deliver cognitive-behavioral interventions through conversational agents, while Replika \cite{replika2026website} is publicly presented as an AI companion application that supports ongoing conversational interaction. While these systems have demonstrated promise, they predominantly focus either on prescriptive cognitive restructuring, AI companionship, or unstructured free-form expression. Few offer a progressive, theory-informed framework that scaffolds emotional development over time.

Reflexion explores this space by proposing a layered, metaphor-driven model for emotional self-reflection. The system integrates emotion detection, cognitive and narrative scaffolding, and metaphorical storytelling to support iterative emotional exploration and cognitive reframing. Table \ref{tab:system_comparison} provides a comparative overview of Reflexion relative to prior systems.

\begin{table*}[t]
\centering
\caption{Comparison of Emotion-Aware AI Systems}
\label{tab:system_comparison}
\begin{tabular}{lccc}
\toprule
\textbf{System} & \textbf{Emotion Signal Use} & \textbf{Narrative Generation} & \textbf{Guided Reflection Layers} \\
\midrule
Woebot & Text-based conversational cues & No & Limited \\
Replika & Memory-supported dialogue cues & Partial (Dialogue) & No explicit layered model \\
Wysa & Text-based mood and CBT cues & No & Limited \\
\textbf{Reflexion (Ours)} & Lightweight transformer soft signals & Yes (Metaphorical) & Explicit four-layer pathway \\
\bottomrule
\end{tabular}
\end{table*}

By explicitly scaffolding emotional self-reflection through structured, layered interactions, Reflexion moves beyond traditional conversational agents toward exploring how AI systems might support emotional articulation, cognitive flexibility, and reflective sense-making.

\section{System Overview}

\subsection{System Architecture: A Modular, Extensible Emotional Reflection Framework}
Reflexion is designed as a modular system to enable scalable, personalized emotional self-reflection. The architecture integrates three core modules, each addressing distinct stages of emotional engagement:

\begin{itemize}
    \item \textbf{Emotion Detection Module}: This module extracts high-level emotional signals from user input using lightweight transformer-based classifiers. Rather than enforcing rigid diagnostic categories, the system surfaces soft emotional cues (e.g., anxiety, self-doubt) to guide reflection, balancing sensitivity with computational efficiency.

    \item \textbf{Guided Reflection Module}: Structured into four progressive layers, this module scaffolds users' emotional processing journeys. It draws primarily from expressive writing \cite{pennebaker1997opening}, cognitive restructuring \cite{beck2021cognitive}, and self-determination theory \cite{ryan2000self} to support self-directed emotional exploration. The final action-planning layer additionally takes loose inspiration from critical consciousness development \cite{freire1970pedagogy}, though this remains an aspirational framing rather than an implemented mechanism.

    \item \textbf{Narrative Generation Module}: To support possible emotional reframing, this module generates metaphorical narratives intended to offer cognitive distance, with the aim of facilitating meaning-making without imposing prescriptive interpretations \cite{white1990narrative}. In the prototype, this module was implemented as a prompt-based generation component: given the user's input, detected emotional cues, and current reflection layer, it produced a short metaphorical narrative.
\end{itemize}

This modular design ensures extensibility, enabling future integration of richer emotion sensing, adaptive prompting, and cross-cultural narrative frameworks.

\paragraph{Backend implementation.}
The emotion detection component was prototyped using a lightweight DistilBERT-based classifier, offering soft-label probability outputs across a set of discrete emotional categories (e.g., joy, sadness, fear, anger). The prototype described in this paper exposes this classifier through a local RESTful interface for emotion signal extraction, and future plans include real-time, context-aware inference and adaptive multi-modal fusion.

\subsection{End-to-End Interaction Flow}
\label{sec:interaction_flow}
\begin{figure}[t]
\centering
\includegraphics[width=0.25\textwidth]{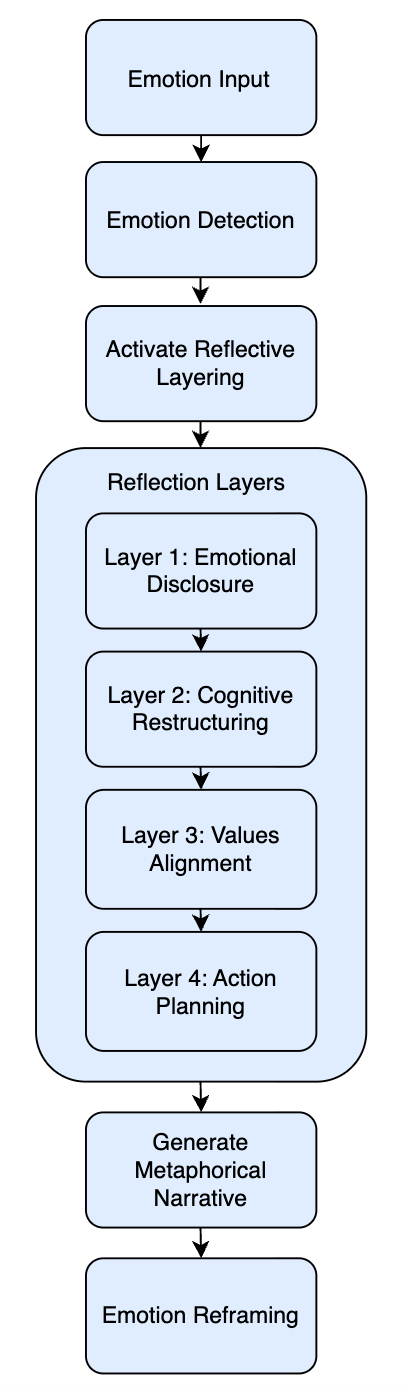}
\caption{Sequential user journey in Reflexion, depicting emotional input, layered reflective scaffolding across four layers, metaphorical narrative generation, and emotional reframing.}
\label{fig:sequential_user_journey}
\end{figure}

Figure~\ref{fig:sequential_user_journey} illustrates the sequential interaction flow implemented in Reflexion, designed to support autonomous emotional self-reflection:

\begin{enumerate}
    \item \textbf{Emotional Input Submission}: Users initiate the process by providing open-ended textual descriptions of their emotional states.
    \item \textbf{Emotion Signal Detection}: The system employs lightweight transformer-based classifiers to infer soft emotional signals.
    \item \textbf{Layered Reflective Scaffolding}: Based on detected signals, users are guided through a multi-layered reflection sequence encompassing emotional disclosure, cognitive restructuring, and values alignment.
    \item \textbf{Metaphorical Narrative Generation}: A metaphorically framed narrative is generated to support cognitive reframing and meaning-making.
    \item \textbf{Cognitive Reframing}: Users are invited to reinterpret their emotional experiences, and the system design allows session-level reflection summaries to support iterative reflection over time.
\end{enumerate}

In addition to this default sequence, the prototype includes a user-initiated \textit{Deep Dive} option. Because everyday contexts rarely require or invite deeper reflection, many users tend to remain at surface-level emotional description. At any point in the flow, the user can press a Deep Dive button to receive prompts that guide reflection into deeper layers on a specific topic, rather than progressing only through the default sequence. Deepening is therefore opt-in and user-triggered, consistent with the autonomy-preserving design philosophy described in Section~\ref{sec:design_philosophy}.

This interaction architecture is designed to progressively scaffold users from raw emotional articulation toward structured cognitive reframing, while aiming to preserve agency and minimize system-imposed interpretation biases.

\subsection{Design Philosophy: Centering Agency and Narrative Sovereignty}
\label{sec:design_philosophy}
Reflexion embraces a human-centered design philosophy, departing from prescriptive chatbot or diagnostic models. Emotions are treated as dynamic, contextually meaningful signals rather than clinical symptoms.

Design choices emphasize:
\begin{itemize}
    \item \textbf{Autonomy Preservation}: Reflection depth and narrative engagement remain user-controlled at every stage; the Deep Dive option makes deeper reflection an explicit user choice rather than a system-imposed progression.
    \item \textbf{Narrative Agency}: Users are invited to reinterpret emotions through self-authored metaphorical lenses, avoiding pre-scripted emotional solutions.
    \item \textbf{Psychological Safety}: The system deliberately avoids coercive nudging or reductive emotional labeling, supporting user control and non-prescriptive reflection.
\end{itemize}

By centering user agency, this design aims to support trust, emotional safety, and sustainable engagement.

\subsection{Critical Design Tradeoffs}
\textbf{Balancing Structure and Flexibility}: While reflection scaffolding is intended to improve emotional articulation, overly rigid pathways risk reducing autonomy. Reflexion mitigates this by allowing optional engagement with each layer, supporting self-determined pacing.

\textbf{Navigating Emotional Complexity vs. Model Simplicity}: Detecting nuanced emotional blends requires rich modeling, but overfitting can harm generalizability. Reflexion opts for interpretable, lightweight emotion detectors as soft guides rather than rigid classifiers.

\textbf{Employing Metaphorical Reframing Safely}: A future metaphor-selection mechanism should prioritize emotional neutrality and adaptive resonance, minimizing unintended cognitive dissonance across diverse users \cite{white1990narrative}.

\textbf{Cultural Generality vs. Localized Resonance}: Emotional norms and metaphorical interpretations vary across cultures \cite{mesquita2014cultural}. The prototype described here uses culturally neutral metaphors but is architected to support future localized narrative adaptations.

\subsection{Integration of Layers: Supporting Progressive Reflection}
Reflexion's layered reflection model is loosely modeled on natural emotional development trajectories. Each layer introduces a different reflective function:
\begin{itemize}
    \item \textbf{Layer 1}: Surface expression (emotional disclosure)
    \item \textbf{Layer 2}: Cognitive restructuring (reframing emotional meaning)
    \item \textbf{Layer 3}: Values alignment (connecting emotions to intrinsic motivations)
    \item \textbf{Layer 4}: Empowered action planning (narrative transformation into agency)
\end{itemize}

This progressive integration model differentiates Reflexion from reactive dialogue systems, positioning it as a prototype for structured, self-directed emotional reflection. Unlike prior AI systems focused on static sentiment analysis \cite{clavel2016sentiment} or cognitive-behavioral dialogue \cite{fitzpatrick2017delivering}, Reflexion organizes reflection as a dynamic, user-led process, offering one design direction for AI-supported emotional literacy tools.

\section{Informal Prototype Review}

This section reports an informal design exploration of the Reflexion prototype. Rather than a controlled empirical study, a small, multi-phase prototype review was conducted to gather early feedback on the system's perceived emotional scaffolding, usability, and theoretical alignment. All impressions reported here are qualitative and self-reported, and should be read as design feedback rather than validated results.

\subsection{Reviewer Recruitment and Profile}

The prototype was informally reviewed by 12 individuals with varied backgrounds, including psychology, computer science, software engineering, education, counseling, and mental health advocacy. These reviewers were invited through the author's academic and professional networks. The process was intended as early design feedback rather than formal participant recruitment.

Reviewers were shown or invited to explore the prototype's Deep Dive reflection pathway (Section~\ref{sec:interaction_flow}).

\subsection{Feedback Procedure}

Reviewers were introduced to the Reflexion prototype and its core interaction design, including emotional input, layered reflective prompts, and metaphorical narrative generation. Depending on access and availability, reviewers either interacted with the prototype directly or reviewed the system workflow and design materials.

Afterward, reviewers provided informal feedback through discussion and written reflections. The feedback focused on perceived emotional scaffolding, usability, narrative resonance, and theoretical alignment. An informal feedback questionnaire and written reflections were collected, but no validated instrument or standardized session protocol was used. Qualitative feedback was collected for design purposes.

\subsection{Emotional Articulation and Granularity}

Several reviewers described Reflexion's layered structure as supporting gradual emotional openness, moving from initial surface-level expression toward deeper emotional exploration. They also described the layered prompts as making it easier to articulate nuanced emotional states during the review session.

Some reviewers described the prompts as helping them think in more nuanced emotional terms, moving beyond broad labels such as ``sad'' or ``angry'' toward more blended descriptions such as ``anxious but hopeful'' or ``frustrated yet determined.''

Several reviewers also described feeling ``more organized'' or ``more in control'' of their emotional experiences after the session. These informal impressions point to possible directions for further design work on Reflexion's layered reflection model, but do not constitute validation.

\subsection{Usability and Flexibility}

Reviewers commented on perceived usability during post-session discussions. Their impressions were generally favorable, but no scores are reported here, as this was an informal review using non-validated feedback questions rather than a standardized usability instrument.

Qualitative feedback highlighted two recurring design impressions:
\begin{itemize}
    \item \textbf{Flexibility:} Some reviewers appreciated the ability to customize their reflection depth and pacing, which they felt reduced emotional pressure.
    \item \textbf{Narrative Resonance:} Some reviewers found metaphorical narratives easier to engage with compared to direct cognitive prompts.
\end{itemize}

These impressions suggest that perceived psychological safety may be a useful dimension to examine in future evaluations of Reflexion, particularly in relation to sustained engagement in affective applications.

\subsection{Perceived Reframing and Autonomy}

Some reviewers described the metaphorical storytelling module as helpful for reframing stressful emotional experiences, reporting that they felt invited to reinterpret negative emotions. Several reviewers also emphasized Reflexion's autonomy-supportive design: unlike diagnostic-oriented tools, the open-ended prompts encouraged voluntary engagement. Reviewers generally described the system as nonjudgmental and autonomy-supportive, noting that the layered structure made emotional exploration feel gradual rather than overwhelming.

After engaging with Reflexion's Deep Dive pathway, several reviewers described emotional reframing as feeling more approachable, and some who explored the deeper reflection layers described these prompts as especially useful for reframing emotionally difficult situations. Some reviewers reported that the prompts helped them reinterpret stressful situations in less self-blaming ways. For example, one type of reflection involved seeing work-related frustration not as personal failure, but as a signal of misalignment with personal values.

\subsection{Summary and Forward Pointer}

Taken together, these informal impressions are encouraging as early design signals but remain unvalidated. Section~\ref{sec:limitations} consolidates the limitations of this review and outlines the formal evaluation that would be required to substantiate any of these impressions as effects.

\section{Discussion}

\subsection{Layered Reflection and Emotional Literacy}

The informal feedback suggested that reviewers perceived the layered structure as helpful for articulating emotional experiences. Several reviewers described Reflexion's progressive structure, combining free expression, cognitive reframing, and narrative generation, as supporting deeper articulation compared with unstructured journaling tools. These impressions should be treated as design signals rather than evidence of effect. If borne out under formal evaluation, this direction could be examined through constructs from scaffolding theory \cite{wood1976scaffolding}, expressive writing research \cite{pennebaker1997opening}, and narrative approaches to meaning-making \cite{bruner1990acts,white1990narrative}.

The increased emotional granularity that some reviewers described, moving from broad labels toward blended descriptions, is of particular interest as a target for future measurement: finer emotional differentiation has been linked in the literature to stronger emotion regulation and more adaptive coping \cite{barrett2001differentiation,kashdan2015unpacking}, and the sense of being ``more organized'' that reviewers reported is the kind of construct addressed by narrative identity research \cite{adler2012narrative}. These are framed here as hypotheses for evaluation, not as findings supported by the present review.

Similarly, the reframing impressions reviewers reported map onto established constructs worth testing directly: narrative therapy principles \cite{white1990narrative}, the autonomy-supportive engagement described by Self-Determination Theory \cite{ryan2000self,ryan2017selfdetermination}, and cognitive reappraisal, which has been associated with more adaptive emotion regulation \cite{gross2003individual}. Unlike conventional chatbot paradigms that emphasize task efficiency or prescriptive wellness advice, Reflexion centers user autonomy and emotional agency, and reviewers described the structured, layered format as useful for self-directed emotional processing.

\subsection{Psychological Safety and Cultural Considerations}

Several reviewers described the prototype as nonjudgmental and low-pressure, suggesting that psychological safety may be an important design consideration for future versions. By avoiding diagnostic labeling and offering non-prescriptive, metaphor-driven prompts, the prototype appeared to support this perception during the review process. This direction is consistent with best practices in trust calibration and emotional validation in human-AI systems \cite{lee2004trust}.

Maintaining narrative agency was particularly valued by reviewers, echoing broader findings that autonomy-supportive interactions increase emotional engagement and intrinsic motivation \cite{ryan2000self}. Several reviewers described Reflexion as ``nonjudgmental'' and ``inviting,'' highlighting the importance of minimizing coercive nudges in emotionally charged contexts \cite{mcstay2020emotional}.

However, cultural variability presents an important frontier. Emotional expression norms differ widely across societies \cite{mesquita2014emotionsociety}, and platforms designed within Western emotional paradigms risk inadvertently alienating non-Western users. Emotional metaphors, central to Reflexion's narrative approach, are also culturally mediated; metaphors intuitive in one cultural context may lack resonance, or even cause confusion, in another \cite{kovecses2004cultural}.

To address this, future work would need to integrate culturally adaptive narrative frameworks and participatory co-design methods involving diverse user groups \cite{stephanidis2019hci}. Psychological safety, therefore, should not be treated as a static system feature but as a dynamic, culturally contextualized experience.

\subsection{Ethical Considerations and Future Research}

As emotionally adaptive AI systems gain traction, ethical considerations must be at the forefront of design. Reflexion's emphasis on user agency, minimal emotional categorization, and non-intrusive guidance reflects emerging ethical standards for emotionally intelligent systems.

Nonetheless, emotional data inherently carries heightened risks concerning privacy, consent, and interpretability \cite{mittelstadt2019ethicalai}. Future development would need to embed transparent data handling practices, user-controlled emotional histories, and explainable emotional inference mechanisms \cite{vilone2021explainable}. Special care is needed to prevent emotional nudging or manipulation, a concern increasingly recognized in the ethics of persuasive AI \cite{mcstay2020emotional}. Reflexion is not designed as a diagnostic, therapeutic, or crisis-support system, and the present prototype should not be interpreted as a substitute for professional mental health care.

Cultural ethics add further complexity: what constitutes emotionally supportive interaction varies across societies \cite{mesquita2014emotionsociety}. Co-design approaches involving cross-cultural users would be vital to ensuring ethical scalability across global contexts.

Finally, the preliminary impressions reported here are not a basis for efficacy claims. Longitudinal, IRB-approved studies would be essential to assess any sustained impact on emotional resilience, self-regulation, and well-being. Interdisciplinary collaboration, bridging affective computing, clinical psychology, educational technology, and AI ethics, would be critical for evolving any successor system responsibly \cite{calvo2014positive}.

\subsection{Limitations and Future Work}
\label{sec:limitations}

Several limitations constrain the scope of this prototype and its informal review.

\textbf{1. No Formal Study and Limited Generalizability:}
This work reports an informal, non-IRB design exploration with 12 reviewers. Although an author-designed feedback questionnaire was used, the absence of a controlled design, validated instruments, and a standardized protocol means the observations should be read as design feedback rather than evidence of effect, and cannot be generalized. Any future formal study should involve a larger and more diverse cohort, including variations in age, cultural backgrounds, and emotional literacy, along with longitudinal follow-up to evaluate sustained emotional change rather than immediate impressions.

\textbf{2. Cultural Variability and Emotional Expression:}
Although Reflexion was designed with a neutral emotional framework, cultural differences in emotional expression and interpretation may affect engagement. Future iterations should consider cross-cultural variations and adapt narrative metaphors through localized co-design and user testing \cite{mesquita2014emotionsociety}.

\textbf{3. Backend Integration for Real-Time Emotion Detection:}
The prototype's emotion detection relies on a pre-trained lightweight classifier, which can be inaccurate for mixed or ambiguous emotional input. A future direction is real-time, context-aware emotion detection that accounts for fluctuating emotional states \cite{wu2025multimodal}.

\textbf{4. Potential Emotional Nudging and Ethical Considerations:}
As with all affective computing systems, there is an inherent risk of emotional nudging or manipulation. While Reflexion's design prioritizes autonomy and non-intrusiveness, stringent safeguards remain necessary. Future research should refine the ethical frameworks around emotional data use, data security, and user privacy \cite{mcstay2018emotionalai}.

\textbf{5. Underdeveloped Critical-Consciousness Layer:}
The action-planning layer draws loose inspiration from critical consciousness development \cite{freire1970pedagogy}, but the current prototype does not instantiate the sociopolitical and collective dimensions central to that framework. Realizing this connection, rather than invoking it, is left to future work.

\textbf{6. Technological Scaling for Broader Deployment:}
For any successor system to reach educational or structured self-reflection contexts, substantial backend scaling and privacy-preserving mechanisms would be required, alongside formal evaluation.

\section{Conclusion}

This manuscript documents Reflexion as an early prototype that drew on affective computing, narrative psychology, and emotional literacy education to explore a human-centered system for AI-supported self-reflection. By scaffolding emotional development through layered reflective pathways and metaphorical storytelling, this version of Reflexion aimed to move beyond prescriptive wellness models toward autonomous emotional exploration.

An informal prototype review produced preliminary design feedback regarding perceived emotional articulation, metaphorical reframing, and psychological safety. Some reviewers responded positively to the layered structure, metaphorical framing, and autonomy-supportive design. Because the review was informal and non-IRB, the reported impressions should be treated as design signals rather than evidence of psychological efficacy.

\section*{Acknowledgments}

The author thanks the reviewers who shared their feedback during the informal prototype review.


\end{document}